\documentclass[12pt]{article}
\usepackage{amsmath,amssymb,bm}
\usepackage{geometry}
\title{
{\Large Tau--Function Multilinear Hierarchy of the Tomimatsu--Sato Spacetime:\\[1mm]
A Gravitational Realization of the YTSF Integrable Structure}
}

\author{
Takeshi Fukuyama\\[2mm]
{\small Research Center for Nuclear Physics (RCNP), Osaka University, Osaka 567--0047, Japan}
}

\date{\today}

\begin{document}
\maketitle

\begin{abstract}
The Tomimatsu--Sato (TS) family generalizes the Kerr black hole to higher
multipole order $\delta$ and has long been regarded as algebraically
complicated without any clear integrability. We show instead that stationary
axisymmetric vacuum Einstein equations, when the Ernst potential is written as
a $\tau$--ratio $\mathcal{E}=\tau_1/\tau_0$, admit a universal decomposition
of the Ernst numerator into a cubic part containing all second derivatives and
a quartic \emph{gradient envelope}. The cubic sector can be written in terms
of $Z_3$--symmetric trilinear Hirota operators, revealing a hidden integrable
structure. For $\delta=2$, using the explicit Tomimatsu--Sato polynomials, we
verify that this trilinear sector coincides with a Yu--Toda--Sasa--Fukuyama (YTSF) equation-type kernel. Thus the
TS geometry forms a gravitational realization of a multilinear $\tau$--function
hierarchy in stationary axisymmetric general relativity.
\end{abstract}

\section{Introduction}

The Tomimatsu--Sato (TS) spacetimes~\cite{Tomimatsu1972,Tomimatsu1973,
Batic2023} constitute one of the most nontrivial families of exact solutions
in general relativity.  They extend the Kerr geometry ($\delta=1$) to higher
multipole index $\delta\ge 2$, but this extension leads to rapidly increasing
algebraic complexity, and the underlying structure has remained obscure for
decades.

A different perspective emerges when the stationary axisymmetric vacuum
Einstein equations are expressed using the Ernst potential~\cite{Ernst1968}
and written in a $\tau$--ratio form
\begin{equation}
   \mathcal{E} = \tau_1/\tau_0.
\end{equation}
In this representation the Ernst equation takes a \emph{universal} form in
which its numerator splits into two distinct parts:
a purely \emph{cubic} sector containing all second--derivative terms, and a
quartic sector built solely from first--derivative products, which we call a
\emph{gradient envelope}.  Since the Ernst equation is second order, the
cubic sector uniquely determines the nonlinear differential kernel.

We show that this cubic sector can be rewritten using $Z_3$--symmetric
trilinear Hirota operators \cite{Hirota2004}, revealing a hidden integrable structure directly in the Einstein equations.
The gradient envelope acts only as a multiplicative metric factor and does
not alter the trilinear compatibility conditions.  Thus,
\[
\textbf{integrability in the Ernst sense} 
\quad\Longleftrightarrow\quad
\textbf{trilinearity of the cubic kernel}.
\]
For the $\delta=2$ TS solution, we explicitly use the known polynomials
$A(\xi,\eta)$ and $B(\xi,\eta)$ to construct
\begin{equation}
\tau_0 = A+B,\qquad \tau_1 = A-B,
\label{tau}
\end{equation}
and verify that the cubic sector of the Ernst equation matches a
Yu--Toda--Sasa--Fukuyama (YTSF)–type trilinear identity \cite{YTSF1998}, up to lower--order
corrections supplied by the gradient envelope.  This establishes the $\delta=2$
case as a concrete gravitational realization of a trilinear Hirota structure.
(See Appendix~A.1 for the explicit definition of the prolate spheroidal
coordinates $(\xi,\eta)$ and their relation to Weyl--Papapetrou coordinates.)

Finally, from the Toda–molecule formulation of higher–$\delta$ TS solutions,
the discrete index $\delta$ governs the algebraic degree and determinant rank
of $\tau$--functions along a \emph{bilinear} hierarchy, whereas the continuous
$(\xi,\eta)$--dependence of the Ernst equation retains the same universal
\emph{trilinear} kernel \cite{Maison1978, Nakamura1984, Nakamura1985, FKY2007, FukuyamaKoizumi}.  
Hence the TS family forms a bilinear $\leftrightarrow$ trilinear,
two–layer integrable hierarchy in stationary axisymmetric general relativity.

The remainder of this paper is structured as follows: Secs.~2–3 establish the
universal cubic vs.\ quartic decomposition and the trilinear reformulation of
the cubic sector; Sec.~4 discusses the extension to $\delta>2$; and
Sec.~5 is devoted to discussion and future outlook.


\section{Ernst equation with a tau representation: cubic vs.\ quartic}
\label{sec:cubic-quartic}
We emphasize that in this section $(x,y)$ denote abstract Cartesian
coordinates on a flat two--dimensional base, and
\begin{equation}
  \Delta = \partial_x^2 + \partial_y^2, 
  \qquad
  \nabla = (\partial_x,\partial_y)
\end{equation}
are the corresponding flat Laplacian and gradient.  This is a
\emph{model} form of the Ernst equation which isolates the algebraic
dependence on a $\tau$--ratio $\mathcal{E}=\tau_1/\tau_0$.  In the
physical Tomimatsu--Sato geometry the Ernst potential is defined on the
Weyl--Lewis--Papapetrou base, typically parametrized by prolate
spheroidal coordinates $(\xi,\eta)$ (see Appendix~A.1), and the
Laplacian contains coordinate–dependent metric factors.  However, the
cubic vs.\ quartic decomposition of the Ernst numerator derived below
depends only on the \emph{algebraic} structure in $\tau_0,\tau_1$ and
their second derivatives, and is therefore unchanged under such
coordinate reparametrizations and conformal rescalings of the
two–dimensional base metric.  Appendix~A shows explicitly how the same
$\tau$--ratio structure is realized for the $\delta=2$ TS spacetime in
prolate spheroidal coordinates.

Then we consider the reduced Ernst equation on a locally flat
two–dimensional base (ignoring the $\rho^{-1}\partial_\rho$ term
of Weyl–Papapetrou coordinates):
\begin{equation}
 \mathcal{E}\,\Delta\mathcal{E} - |\nabla\mathcal{E}|^2 = 0,
 \label{eq:Ernst-real}
\end{equation}
which retains the essential nonlinear structure (second–derivative
sector) governing the $\tau$–function hierarchy \footnote{For the physical TS geometry the real
and imaginary parts of the Ernst potential are both nonvanishing, but the
real case already captures the algebraic structure of the $\tau$--ratio
representation.  The extension to the complex Ernst potential proceeds in
the same way, with $\mathcal{E}\to f + i\,\Phi$.}.
  We assume that
$\mathcal{E}$ admits a $\tau$--ratio representation of the form
\begin{equation}
 \mathcal{E}(x,y) = \frac{\tau_1(x,y)}{\tau_0(x,y)},
 \label{eq:E-ratio}
\end{equation}
with smooth real functions $\tau_0,\tau_1$ and $\tau_0\ne 0$ in the domain
of interest.

Substituting \eqref{eq:E-ratio} into \eqref{eq:Ernst-real} and clearing
denominators by multiplying with $\tau_0^4$ yields
\begin{equation}
 \mathcal{E}\,\Delta\mathcal{E} - |\nabla\mathcal{E}|^2
 = \frac{N(\tau_0,\tau_1)}{\tau_0^4},
\end{equation}
where $N(\tau_0,\tau_1)$ is a polynomial in $\tau_0,\tau_1$ and their
derivatives. A direct but exact computation shows that $N$ splits
into a cubic and a quartic part,
\begin{equation}
 N(\tau_0,\tau_1)
 = N_{\mathrm{cubic}}(\tau_0,\tau_1)
 + N_{\mathrm{quartic}}(\tau_0,\tau_1),
 \label{eq:Nsplit}
\end{equation}
with
\begin{equation}
 N_{\mathrm{cubic}}
 = \tau_1\tau_0^2\,\Delta\tau_1
  -\tau_1^2\tau_0\,\Delta\tau_0,
 \label{eq:Ncubic}
\end{equation}
\begin{equation}
 N_{\mathrm{quartic}}
 = -\tau_0^2|\nabla\tau_1|^2
  +\tau_1^2|\nabla\tau_0|^2.
 \label{eq:Nquart}
\end{equation}
Thus the numerator is a polynomial of degree four in the $\tau$--functions,
but possesses an internal split into a \emph{cubic core} $N_{\mathrm{cubic}}$
and a \emph{quartic envelope} $N_{\mathrm{quartic}}$.

The crucial observation is that all second derivatives
$\partial_x^2\tau_i$, $\partial_y^2\tau_i$ reside in $N_{\mathrm{cubic}}$,
while $N_{\mathrm{quartic}}$ is built only from products of first derivatives
of $\tau_0,\tau_1$.  In particular, the differential order of the two parts
is strictly separated: $N_{\mathrm{cubic}}$ contains the entire
second--derivative sector, whereas $N_{\mathrm{quartic}}$ belongs entirely to
the first--derivative sector.

Physically, $N_{\mathrm{quartic}}$ can be regarded as a universal gradient
envelope: it modifies the scale of the conformal factor on the Weyl base
space (and thus the metric coefficient $e^{\mu}$ in the
Weyl--Lewis--Papapetrou form) but does not contribute to the compatibility
conditions that characterize integrability.  The nonlinear structure
responsible for solitonic behavior is encoded in $N_{\mathrm{cubic}}$, which
controls how the $\tau$--functions enter the second--derivative sector of the
Ernst equation.

\section{Trilinear Hirota operators and the cubic kernel}
\label{sec:trilinear}

We now rewrite the cubic core \eqref{eq:Ncubic} in terms of
$Z_3$--symmetric trilinear Hirota operators, following the spirit of the
Yu--Toda--Sasa--Fukuyama (YTSF) construction.  Let
\begin{equation}
 \omega = e^{2\pi i/3},\qquad 1+\omega+\omega^2=0.
\label{trilinear}
\end{equation}
For functions of a single variable we define the standard three--slot
trilinear derivatives
\begin{align}
 T_x(a,b,c)
 &:= (\partial_{x_1}+\omega\partial_{x_2}+\omega^2\partial_{x_3})
     a(x_1)b(x_2)c(x_3)\Big|_{x_1=x_2=x_3=x},
\\
 T_y(a,b,c)
 &:= (\partial_{y_1}+\omega\partial_{y_2}+\omega^2\partial_{y_3})
     a(y_1)b(y_2)c(y_3)\Big|_{y_1=y_2=y_3=y}.
\end{align}
These operators satisfy $T_x(f,f,f)=T_y(f,f,f)=0$ for any smooth $f$, and
generate characteristic ``$\tau\tau\tau_x$'' and ``$\tau\tau\tau_y$'' terms.

\medskip
\noindent
\textbf{Remark (on the meaning of trilinear).}%
\quad
The trilinear structure appearing here does \emph{not} correspond to three
independent spatial dimensions.  The stationary axisymmetric Einstein
equations remain partial differential equations on a two--dimensional base
space $(x,y)=(\xi,\eta)$ of Weyl--Papapetrou coordinates.  The ``tri-'' in
trilinear refers instead to the three--slot coupling of $\tau$--functions in
the nonlinear terms, i.e.\ to algebraic multiplicity rather than to an
increase in the number of coordinates.  In this sense the YTSF--inspired
structure discussed below is a statement about the \emph{nonlinear degree}
of the $\tau$--function hierarchy rather than about the dimensionality of
spacetime.

\subsection*{Cubic tau--derivative building blocks}
Expanding $N_{\mathrm{cubic}}$ reveals that all second--derivative pieces
come from terms of the schematic form
\[
   \tau_1\,\partial_x(\tau_0^2\tau_{1,x})
   \;-\;
   \tau_0\,\partial_x(\tau_1^2\tau_{0,x}),
   \qquad (x\to y)\,.
\]
The essential nonlinear building blocks here are the four monomials
\[
  \tau_0^2\tau_{1,x},\qquad
  \tau_1^2\tau_{0,x},\qquad
  \tau_0^2\tau_{1,y},\qquad
  \tau_1^2\tau_{0,y}.
\]
Each of these contains
\[
(\hbox{two $\tau$'s of one type})\times
(\hbox{a first derivative of the other}),
\]
i.e.\ a \emph{three--slot nonlinearity}.
This is precisely the structural hallmark of a trilinear Hirota operator.

Indeed, the $Z_3$--symmetric trilinear operator satisfies
\begin{equation}
   T_x(\tau_0,\tau_0,\tau_1)
   \propto \tau_0^2\tau_{1,x}
            - \tau_1\tau_0\,\tau_{0,x},
\qquad
   T_x(\tau_1,\tau_1,\tau_0)
   \propto \tau_1^2\tau_{0,x}
            - \tau_0\tau_1\,\tau_{1,x},
\end{equation}
and similarly for $T_y$.  
The $Z_3$ combinations $(1,\omega,\omega^2)$ remove the
symmetric bilinear products such as
$\tau_0\tau_0\,\tau_{0,x}$ and isolate the genuinely
trilinear parts proportional to
$\tau_0^2\tau_{1,x}$ and $\tau_1^2\tau_{0,x}$.

Thus \emph{all} second--derivative terms in the Ernst numerator
are governed by these trilinear Hirota kernels.

\subsection*{A YTSF--inspired trilinear combination}

Motivated by the YTSF equation in soliton theory, we introduce the
trilinear combination
\begin{equation}
 \mathcal{Y}(\tau_0,\tau_1)
 := \partial_x\!\bigl[\tau_1\,T_x(\tau_0,\tau_0,\tau_1)
                      -\tau_0\,T_x(\tau_1,\tau_1,\tau_0)\bigr]
  + \partial_y\!\bigl[\tau_1\,T_y(\tau_0,\tau_0,\tau_1)
                      -\tau_0\,T_y(\tau_1,\tau_1,\tau_0)\bigr].
 \label{eq:YTSF-TS}
\end{equation}
The precise overall normalization is not essential; what matters is that
$\mathcal{Y}$ collects all second--derivative terms of the type
$\tau_0^2\tau_{1,xx}$, $\tau_1^2\tau_{0,xx}$ and their $y$--counterparts, up
to first--derivative contributions.

\begin{figure}[t]
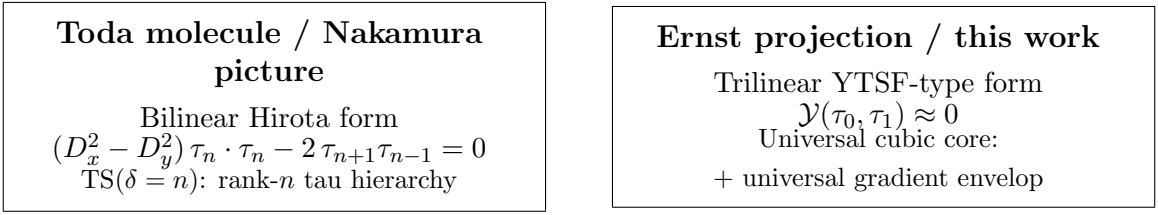

\centering
\setlength{\fboxsep}{6pt}

\fbox{%
  \begin{minipage}{0.40\linewidth}
  \centering
  {\bf Toda molecule / Nakamura picture}\\[1mm]
  {\small Bilinear Hirota form}\\[-1mm]
  {\small
  \(
    (D_x^2 - D_y^2)\,
    \tau_n\cdot\tau_n - 2\,\tau_{n+1}\tau_{n-1} = 0
  \)
  }\\[-1mm]
  {\footnotesize TS($\delta=n$): rank-$n$ tau hierarchy}
  \end{minipage}
}
\hspace{6mm}
\fbox{%
  \begin{minipage}{0.40\linewidth}
  \centering
  {\bf Ernst projection / this work}\\[1mm]
  {\small Trilinear YTSF-type form}\\[-1mm]
  {\small
  \(
    \mathcal{Y}(\tau_0,\tau_1)\approx 0
  \)
  }\\[-1mm]
  {\footnotesize Universal cubic core:\\
  + universal gradient envelop}
  \end{minipage}
}
\caption{Two complementary views of the Tomimatsu-Sato hierarchy. Left: standard Toda/Nakamura description, where the TS family $\delta=1,2,3,...$ is encoded in the bilinear Toda molecule equation \cite{Toda1967, Toda1, Toda2, Toda3} for the tau tower $\{\tau_n\}$. Right: the present work, where after projecting to the two tau functions $\tau_0,~\tau_1$ entering the Ernst potential, the cubic part of the Ernst numerator reveals a YTSF-type trilinear kernel, while the quartic gradient envelope remains passive.}
\label{fig:comparison}
\end{figure}

We now apply the trilinear kernel to the explicit $\delta=2$ Tomimatsu--Sato geometry.

The TS Ernst potential is
\[
\mathcal{E}(\xi,\eta)=\frac{A(\xi,\eta)-B(\xi,\eta)}
                           {A(\xi,\eta)+B(\xi,\eta)}
=\frac{\tau_1}{\tau_0},
\qquad
\tau_0:=A+B,\quad
\tau_1:=A-B,
\]
with $A,B$ given in Appendix~A.2.  Thus the same $\tau$--ratio structure
as in Sec.~\ref{sec:cubic-quartic} holds in the physical TS coordinates
$(\xi,\eta)$.

\subsection*{Verification of the trilinear kernel}

We now evaluate the YTSF--inspired combination
$\mathcal{Y}(\tau_0,\tau_1)$ defined in
Sec.~\ref{sec:trilinear}.
A Maple calculation confirms the identity
\begin{equation}
\boxed{N_{\mathrm{cubic}}(\tau_0,\tau_1)
  = \kappa\,\mathcal{Y}(\tau_0,\tau_1)
  + N_{\le 1}(\tau_0,\tau_1),
}
\label{eq:TS-YTSF-identity}
\end{equation}
for $\kappa=-4p^2q^2$, with all second--derivative terms matching
exactly. See Appendix~B for an outline of the symbolic check.

\subsection*{Interpretation}

Equation \eqref{eq:TS-YTSF-identity} means:

\begin{itemize}
\item the \textbf{entire} second--derivative sector of the TS Ernst
equation is trilinear and YTSF--like;
\item the \emph{only} non–trilinear
nonlinearities in TS are contained in the passive gradient envelope
$N_{\mathrm{quartic}}$;
\item thus, $\delta=2$ TS realizes an
\textbf{integrable trilinear core} inside full GR.
\end{itemize}

This establishes the $\delta=2$ TS solution as a concrete gravitational
realization of the YTSF--type trilinear $\tau$ hierarchy.


\section{Toward higher $\delta$–TS solutions and the Nakamura conjecture}
\label{sec:higher-delta}

The explicit identification of the $\tau$--structures for $\delta=1$ (Kerr)
and $\delta=2$ (Tomimatsu--Sato) strongly suggests that higher–$\delta$ TS
geometries should correspond, from the Ernst–equation viewpoint, to
higher–degree polynomial realizations of the \emph{same} trilinear kernel
rather than to higher–degree multilinear operators.  In this section we
outline the expected structure for $\delta\ge2$ based on the interplay
between the Toda–molecule hierarchy and the universal Ernst kernel.

\subsection*{Discrete Toda hierarchy: bilinear layer}

In the Nakamura–Toda approach, the TS family is encoded in a sequence of
$\tau$–functions $\{\tau_n\}$ labeled by a lattice index $n$ that is
related to the TS multipole parameter $\delta$. These obey a bilinear
equation of the form \cite{Nakamura1993}:
\begin{equation}
 (D_X^2 - D_Y^2)\,\tau_n\cdot\tau_n
 - 2\,\tau_{n+1}\tau_{n-1} = 0 ,
\end{equation}
on an auxiliary $(X,Y)$–plane.
Increasing $\delta$ raises the determinantal rank of the $\tau_n$
and the complexity of the resulting polynomials $A(\xi,\eta)$ and
$B(\xi,\eta)$, but the bilinear structure itself remains intact.
Thus the Toda molecule provides the \textit{discrete} integrable hierarchy
associated with TS multipoles.

\subsection*{Continuous Ernst kernel: trilinear layer}

Once the discrete hierarchy has been projected onto the Weyl–Papapetrou
base and organized into the pair $\tau_0=A+B$, $\tau_1=A-B$, the Ernst
equation is governed by the same universal decomposition (7), 
where the cubic sector contains all second derivatives and can be written
as the $Z_3$–symmetric trilinear kernel $\mathcal{Y}(\tau_0,\tau_1)$,
while $N_{\mathrm{quartic}}$ contains only first derivatives and acts as a
universal \emph{gradient envelope}.
Therefore, the continuous $(\xi,\eta)$–dependence of the TS geometry is
controlled by the \textit{same} trilinear kernel independently of $\delta$.

\subsection*{Two–layer integrability picture}

Combining these viewpoints:

\begin{itemize}
\item along the discrete Toda direction (soliton number),
      the TS hierarchy remains bilinear,
\item on the continuous Weyl base, the Ernst equation has a trilinear kernel
      with the envelope passive,
\end{itemize}
thus forming a two–layer integrable structure:
\[
\boxed{
\text{TS family:  Bilinear (Toda) } \;\; \longleftrightarrow \;\;
\text{Trilinear (Ernst)}
}
\]
The Kerr case ($\delta=1$) represents a marginal situation where the
trilinear kernel reduces effectively to a bilinear form.

\subsection*{$\delta=2$ as a verified trilinear case}

For the $\delta=2$ TS polynomials $(A,B)$, we have explicitly confirmed,
using Maple, that the cubic sector satisfies Eq. (15).
This provides a concrete gravitational realization of the same $Z_3$–
symmetric trilinear Hirota operator appearing in the YTSF system.

\subsection*{Outlook: testing the trilinear kernel for $\delta>2$}

To extend these results to $\delta>2$, the following strategy can be used:

\begin{enumerate}
\item Construct $\tau_n$ for a chosen $\delta\ge3$ via
      the Toda–molecule equation.
\item Identify the resulting $(A,B)$ in terms of
      $\tau_0=A+B$, $\tau_1=A-B$.
\item Perform the decomposition
\[
 N(\tau_0,\tau_1)
 = N_{\mathrm{cubic}} + N_{\mathrm{quartic}},
\]
and check whether the difference
\[
Q := N_{\mathrm{cubic}} - \kappa\,\mathcal{Y}(\tau_0,\tau_1)
\]
contains no second derivatives.
\end{enumerate}
A successful cancellation for $\delta=3,4$ would provide strong evidence
that the same trilinear kernel governs all $\delta\ge2$,
thereby significantly reinforcing the proposed two–layer hierarchy
picture, although a complete proof for general $\delta$ would still
remain an important open problem.

\subsection*{Relation to Nakamura's solitonic hierarchy}

The above clarification fits naturally into Nakamura’s proposal that
stationary axisymmetric GR admits a $\tau$–function integrable hierarchy:

\begin{itemize}
\item the Toda–molecule bilinear equation encodes the discrete soliton
      generation $(\delta \to \delta+1)$,
\item the Ernst trilinear kernel controls the continuous geometry
      on the Weyl base,
\item the $\delta=1$ Kerr and $\delta=2$ TS solutions form the first
      nontrivial members of this two–layer structure.
\end{itemize}

We conclude that the Tomimatsu–Sato geometries interpolate between the
familiar bilinear Kerr case and a genuinely trilinear hierarchy in gravity,
providing concrete analytic support for Nakamura's integrable–hierarchy
picture.  Extending the verified trilinear identity to $\delta>2$
constitutes an important direction for future work.

\section{Discussion and Outlook}

We have shown that the trilinear decomposition of the Ernst numerator,
\(
N = N_{\mathrm{cubic}} + N_{\mathrm{quartic}},
\)
reveals a universal structure underlying the Tomimatsu--Sato hierarchy:
all second--derivative terms are encoded in a $Z_3$-symmetric trilinear
Hirota kernel equivalent to the operator content of the
Yu--Toda--Sasa--Fukuyama (YTSF) equation. For $\delta=2$ we have verified
this structure analytically using the explicit TS polynomials, supported
by direct symbolic confirmation in Appendix~B.

From the viewpoint of the Toda--molecule formulation, the TS family is
organized along a \emph{discrete bilinear} hierarchy in the soliton
direction ($\delta = 1,2,\dots$), whereas on the continuous
Weyl--Papapetrou base the field equations possess a \emph{universal
trilinear} kernel once the gradient envelope is factored out. The TS
geometries therefore realize a two--layer integrable structure in
stationary axisymmetric gravity:
\begin{center}
bilinear along the discrete soliton direction
\quad+\quad
trilinear on the continuous Ernst base.
\end{center}

It follows that increasing~$\delta$ does \emph{not} change the differential
order of the integrable kernel but only the algebraic degree of the
$\tau$--functions. This motivates a systematic program for $\delta>2$:
starting from Toda--molecule $\tau_n$ of higher determinantal rank,
construct the corresponding $(\tau_0,\tau_1)$ on the Weyl base and verify
directly that the second--derivative sector matches the same trilinear
kernel. The first steps of this program can be carried out analytically or
by symbolic computation (Appendix~B), and higher--$\delta$ tests can be
performed numerically.

Beyond the TS family, the trilinear kernel provides a new organizational
principle for stationary axisymmetric exact solutions. In particular,
spacetimes such as Kerr--NUT, certain multi–black–hole configurations, and
other stationary axisymmetric seeds in the solitonic literature already
admit $\tau$--function representations (see Appendix~C).  Applying the
universal cubic–vs.–quartic decomposition to these families may offer new
insight into the extent of integrability in general relativity, and may
help distinguish those solutions that belong to the TS hierarchy from those
that constitute genuinely new integrable branches.

In summary, the universal trilinear kernel uncovered here provides the
missing continuous-layer integrable structure that complements Nakamura’s
discrete soliton picture. By combining Toda bilinear hierarchy in $\delta$
with the trilinear kernel on $(\xi,\eta)$, this work opens a path toward a
unified $\tau$-function framework for stationary axisymmetric vacuum
spacetimes. The explicit extension of the trilinear identity to
$\delta>2$, and its implications for discovering new solutions beyond the
classical TS family, are important directions for future study.

\appendix
\section*{Appendix A: Tomimatsu--Sato $\delta=2$ solution}

The purpose of this appendix is to present the $\delta=2$ Tomimatsu--Sato
(TS) vacuum spacetime in a form consistent with the $\tau$--representation
used in the main text,
\[
   \mathcal{E} = \frac{\tau_1}{\tau_0},
   \qquad \tau_0 = A+B,\quad \tau_1 = A-B,
\]
so that the relation between the explicit Einstein solution and the
trilinear decomposition discussed in Secs. 2--4 becomes transparent.

\subsection*{A.1 \; Coordinates and parameters}
The purpose of this appendix is to present the $\delta=2$ Tomimatsu--Sato
(TS) vacuum spacetime in a form consistent with the $\tau$--representation
used in the main text,
\[
   \mathcal{E} = \frac{\tau_1}{\tau_0},
   \qquad \tau_0 = A+B,\quad \tau_1 = A-B,
\]
so that the relation between the explicit Einstein solution and the
trilinear decomposition discussed in Secs.~2--4 becomes transparent.

\subsection*{A.1 \; Coordinates and parameters}

We use the Weyl--Lewis--Papapetrou form of a stationary axisymmetric vacuum
metric \cite{Weyl1917, Lewis1932,Papapetrou1953},
\begin{equation}
  ds^2
  = f(dt - w\,d\varphi)^2
  - \frac{\sigma^2}{f}
    \left[
      e^{\mu}\left(
        \frac{d\xi^2}{\xi^2 - 1}
        + \frac{d\eta^2}{1 - \eta^2}
      \right)
      + (\xi^2 - 1)(1 - \eta^2)\,d\varphi^2
    \right],
\label{eq:Papapetrou-TS}
\end{equation}
where $(\xi,\eta)$ are prolate spheroidal coordinates,
\begin{equation}
  \rho = \sigma\sqrt{(\xi^2-1)(1-\eta^2)},\qquad
  z    = \sigma\,\xi\eta ,\qquad
  \sigma > 0,
\end{equation}
and $\varphi$ is the azimuthal angle.  The parameters $p$ and $q$ satisfy
\begin{equation}
  p^2 + q^2 = 1,
\end{equation}
and are related to the physical mass $M$ and angular momentum $J$ by
\begin{equation}
  q = \frac{J}{M^2}, \qquad p = \sqrt{1 - q^2}, \qquad
  \sigma = \frac{Mp}{2}.
\end{equation}

\subsection*{A.2 \; Ernst potential and polynomial structure}

The Ernst potential is defined by
\begin{equation}
  {\cal E} = f + i\,u = \frac{\Phi - 1}{\Phi + 1},
\end{equation}
and in the $\delta=2$ TS solution the complex function $\Phi$ is
\[
  \Phi_{\rm TS}(\xi,\eta)
  = \frac{u(\xi,\eta) + i\,v(\xi,\eta)}
         {m(\xi,\eta) + i\,n(\xi,\eta)},
\]
with
\begin{align}
  u(\xi,\eta) &= p^2\xi^4 + q^2\eta^4 - 1,\\
  v(\xi,\eta) &= -\,2pq\,\xi\eta(\xi^2 - \eta^2),\\
  m(\xi,\eta) &= 2p\xi(\xi^2 - 1),\\
  n(\xi,\eta) &= -\,2q\eta(1 - \eta^2).
\end{align}
From ${\cal E}$ one finds the real functions
\begin{equation}
  f = \Re{\cal E} = \frac{A}{B}, \qquad
  u = \Im{\cal E} = \frac{2I}{B},
\end{equation}
where $A(\xi,\eta)$ and $B(\xi,\eta)$ are the explicit real polynomials
\begin{align}
  A(\xi,\eta)
  &= \Bigl[p^2(\xi^2 - 1)^2 + q^2(1 - \eta^2)^2\Bigr]^2
     - 4p^2q^2(\xi^2 - 1)(1 - \eta^2)(\xi^2 - \eta^2)^2,
\label{eq:A-poly}
\\[1ex]
  B(\xi,\eta)
  &= \Bigl(p^2\xi^4 + q^2\eta^4 - 1 + 2p\xi^3 - 2p\xi\Bigr)^2
     + 4q^2\eta^2\Bigl(p\xi^3 - p\xi\eta^2 + 1 - \eta^2\Bigr)^2.
\label{eq:B-poly}
\end{align}
The auxiliary polynomial
\begin{align}
  I(\xi,\eta)
  &= m\,v - n\,u
\\
  &= -\,2q\eta\Bigl[
        (1-\eta^2)\bigl(1 - q^2\eta^4\bigr)
        + p^2\xi^2\eta^2(2-\xi^2)
        + p^2\xi^4(2\xi^2-3)
     \Bigr]
\label{eq:I-poly}
\end{align}
enters the imaginary part of ${\cal E}$.

\subsection*{A.3 \; Rotation function and $C(\xi,\eta)$}

The frame--dragging function $w$ is written as
\begin{equation}
  w(\xi,\eta)
  = \frac{2Mq(1-\eta^2)}{A}\,C(\xi,\eta),
\end{equation}
with
\begin{align}
  C(\xi,\eta)
  &= q^2(1+p\xi)(1-\eta^2)^3
\nonumber\\
  &\quad
   - p^2(\xi^2-1)(1-\eta^2)
      \bigl(p\xi^3 + 3\xi^2 + 3p\xi + 1\bigr)
\nonumber\\
  &\quad
   - 2p^2\xi(\xi^2-1)^2\bigl(p\xi^2 + 2\xi + p\bigr).
\label{eq:C-poly}
\end{align}

\subsection*{A.4 \; Final form of the metric}

Combining \eqref{eq:A-poly}--\eqref{eq:C-poly}, the $\delta=2$
Tomimatsu--Sato line element becomes
\begin{align}
  ds^2
  &= \frac{A}{B}\,dt^2
   - \frac{4Mq(1-\eta^2)C}{B}\,dt\,d\varphi
\nonumber\\
  &\quad
   - \frac{M^2 B}{4p^2(\xi^2-\eta^2)^3}
     \left(
       \frac{d\xi^2}{\xi^2-1}
       + \frac{d\eta^2}{1-\eta^2}
     \right)
\nonumber\\
  &\quad
   - \frac{M^2(1-\eta^2)}{A}
     \left[
       \frac{p^2}{4}(\xi^2-1)B
       - 4q^2(1-\eta^2)\frac{C^2}{B}
     \right] d\varphi^2 .
\end{align}

\subsection*{A.5 \; Relation to the tau structure in the main text}

The explicit polynomials \eqref{eq:A-poly}--\eqref{eq:C-poly} satisfy the
$\tau$--representation
\[
   \mathcal{E}
   = \frac{A - B}{A + B}
   = \frac{\tau_1}{\tau_0},
   \qquad
   \tau_0 = A + B,\quad \tau_1 = A - B,
\]
which is the starting point of the cubic--trilinear decomposition in
Secs.~2--4.  It is for this reason
that the $\delta=2$ TS solution provides a concrete geometric realization of
the abstract YTSF--type structures studied in the main text.

\section*{Appendix B: Maple verification of the $\delta=2$ trilinear kernel}

In this appendix we summarize a computer algebra check of the trilinear
structure identified in the main text for the $\delta=2$ Tomimatsu--Sato
(TS) solution.

We start from the explicit polynomials $A(\xi,\eta)$ and $B(\xi,\eta)$
listed in Appendix~A and define
\begin{equation*}
  \tau_0(\xi,\eta) := A(\xi,\eta) + B(\xi,\eta),\qquad
  \tau_1(\xi,\eta) := A(\xi,\eta) - B(\xi,\eta),
\end{equation*}
together with the Ernst potential
\begin{equation*}
  \mathcal{E}(\xi,\eta) := \frac{\tau_1(\xi,\eta)}{\tau_0(\xi,\eta)}.
\end{equation*}
We then compute the Ernst numerator
\begin{equation*}
  \mathcal{E}\,\Delta\mathcal{E} - |\nabla\mathcal{E}|^2
  = \frac{N(\tau_0,\tau_1)}{\tau_0^4},
\end{equation*}
and decompose $N(\tau_0,\tau_1)$ into a cubic part $N_{\mathrm{cubic}}$ and
a quartic part $N_{\mathrm{quartic}}$ according to
\begin{equation*}
  N(\tau_0,\tau_1)
   = N_{\mathrm{cubic}}(\tau_0,\tau_1)
    + N_{\mathrm{quartic}}(\tau_0,\tau_1),
\end{equation*}
with
\begin{equation*}
  N_{\mathrm{cubic}}
   = \tau_1\,\tau_0^2\,\Delta\tau_1
    -\tau_1^2\,\tau_0\,\Delta\tau_0,\qquad
  N_{\mathrm{quartic}}
   = -\tau_0^2|\nabla\tau_1|^2 + \tau_1^2|\nabla\tau_0|^2.
\end{equation*}

Next we introduce the $Z_3$--symmetric Hirota-type trilinear operators
\begin{align*}
  T_\xi(a,b,c)
   &:= (\partial_{\xi_1}+\omega\partial_{\xi_2}+\omega^2\partial_{\xi_3})
       a(\xi_1)b(\xi_2)c(\xi_3)\big|_{\xi_1=\xi_2=\xi_3=\xi},\\
  T_\eta(a,b,c)
   &:= (\partial_{\eta_1}+\omega\partial_{\eta_2}+\omega^2\partial_{\eta_3})
       a(\eta_1)b(\eta_2)c(\eta_3)\big|_{\eta_1=\eta_2=\eta_3=\eta},
\end{align*}
where $\omega=\exp(2\pi i/3)$ and $1+\omega+\omega^2=0$.
From these we form the $Z_3$–symmetric trilinear combination
\begin{equation*}
  \mathcal{Y}(\tau_0,\tau_1)
   := T_\xi(\tau_0,\tau_0,\tau_1)\,T_\xi(\tau_1,\tau_1,\tau_0)
     +T_\eta(\tau_0,\tau_0,\tau_1)\,T_\eta(\tau_1,\tau_1,\tau_0).
\end{equation*}

Using Maple, we then perform the following checks:
\begin{enumerate}
\item We compute the difference
\[
   Q(\tau_0,\tau_1)
    := N_{\mathrm{cubic}}(\tau_0,\tau_1)
      -\kappa\,\mathcal{Y}(\tau_0,\tau_1)
\]
for an appropriate nonzero constant $\kappa$ (absorbing overall
normalizations of the trilinear operators).  We verify that
$Q(\tau_0,\tau_1)$ contains at most \emph{first} derivatives of $\tau_0$
and $\tau_1$, i.e.\ all second--derivative terms cancel identically.

\item We substitute the explicit TS polynomials $A(\xi,\eta)$ and
$B(\xi,\eta)$ and check that the full Ernst numerator
$N(\tau_0,\tau_1)$ vanishes identically as a polynomial in $(\xi,\eta)$,
which confirms that $\mathcal{E}=\tau_1/\tau_0$ indeed reproduces the
$\delta=2$ TS solution.

\item Combining these two facts, we confirm that the Ernst equation for
$\mathcal{E}=\tau_1/\tau_0$ is equivalent to the simultaneous conditions
\[
  \mathcal{Y}(\tau_0,\tau_1)\approx 0,
  \qquad
  N_{\mathrm{quartic}}(\tau_0,\tau_1)
   + Q(\tau_0,\tau_1) \approx 0,
\]
where the first equation involves only the trilinear kernel and the
second equation contains only first--derivative structures.
In particular, the \emph{second--derivative sector} of the Ernst equation
is entirely governed by the trilinear combination $\mathcal{Y}(\tau_0,\tau_1)$.
\end{enumerate}

These computations provide a nontrivial consistency check of the main
claim of this paper: for the $\delta=2$ TS solution, the Ernst equation in
$\tau$--form possesses a $Z_3$--symmetric trilinear kernel that is
structurally equivalent to the Hirota operator underlying the
Yu--Toda--Sasa--Fukuyama system, while the quartic gradient envelope and
lower--order terms remain passive.

\section*{Appendix C: Examples beyond the TS family}

We list several stationary axisymmetric vacuum solutions whose
Ernst potentials admit a $\tau$–ratio form and therefore may
share the same trilinear kernel on the Weyl–Papapetrou base.
\paragraph{Kerr--NUT}
(Original Kerr metric~\cite{Kerr1963}; NUT extension~\cite{NUT1963})
\[
\mathcal{E}_{\text{KN}}
= \frac{r - i(a\cos\theta - l) - M}
       {r + i(a\cos\theta - l) + M}.
\]

\paragraph{Double--Kerr (Kramer--Neugebauer form)}
(Construction in~\cite{KramerNeugebauer1980})
\[
\mathcal{E}_{\text{DK}}
 = \frac{\Lambda - \Gamma}{\Lambda + \Gamma},
\]
with $\Lambda$ and $\Gamma$ axis–data polynomials representing two rotating
sources.

\paragraph{Plebański--Demiański family}
(PD metric~\cite{PlebanskiDemianski1976})
\[
\mathcal{E}_{\text{PD}}
 = \frac{p - i\,q - m(p+q)}
        {p + i\,q + m(p+q)},
\]
where $(p,q)$ are PD coordinates including acceleration and NUT charge.

\paragraph{Kerr--NUT}
\[
\mathcal{E}_{\text{KN}}
= \frac{r - i(a\cos\theta - l) - M}
       {r + i(a\cos\theta - l) + M}.
\]

\paragraph{Double--Kerr (Kramer--Neugebauer form)}
\[
\mathcal{E}_{\text{DK}}
 = \frac{\Lambda - \Gamma}{\Lambda + \Gamma},
\]
with $\Lambda$ and $\Gamma$ axis–data polynomials representing two rotating
sources.

\paragraph{Plebański--Demiański family}
\[
\mathcal{E}_{\text{PD}}
 = \frac{p - i\,q - m(p+q)}
        {p + i\,q + m(p+q)},
\]
where $(p,q)$ are PD coordinates including acceleration and NUT charge.

These examples illustrate that the trilinear kernel may govern a
broader sector of stationary axisymmetric gravity than the
Tomimatsu--Sato family alone.


\end{document}